# The Potential-Vortex Theory of the Electromagnetic Field

Tomilin A.K.[*]

*D. Serikbayev East-Kazakhstan State Technical University
Ust-Kamenogorsk, Kazakhstan*

## Abstract

Maxwell-Lorenz theory describes only vortex electromagnetic processes. Potential component of the magnetic field is usually excluded by the introduction of mathematical terms: Coulomb and Lorenz gauges. Proposed approach to the construction of the four-dimensional electrodynamics based on the total (four-dimensional) field theory takes into account both vortex and potential components of its characteristics. It is shown that potential components of the electromagnetic field have physical content. System of modified (generalized) Maxwell equations is written. With their help contradictions usually appearing while describing the distribution of electromagnetic waves, are eliminated.

Works of other authors obtained similar results are presented and analyzed.

---
[*] e-mail: AKTomilin@gmail.com

**Introduction**

Typically, the study of electrodynamics follows the historical path: first, separate electrical and magnetic phenomena are considered, and then transfer to forming of the electromagnetic field conception. The top of the electrodynamic theory are Maxwell's equations, of which, as a consequence, wave equations are derived.

Let us turn to the analogy between mechanics and electrodynamics. Modern physicists generally believe that a mechanistic description is fundamentally limited and unsuitable for the study of electromagnetic processes. This point of view is seriously rejected in the substantial article of P.A. Zhilin [1]. In particular it is shown that the analogy between the theory of elasticity and electrodynamics allows us to find out fundamental limitations of the theory based on Maxwell's equations.

In the theory of elasticity (and generally in continuum mechanics), general field approach is used in full: due to the motion of particles medium is deformed, potential and curly currents appear, and vice versa, particles are entrained by movement of the medium. This can be clearly seen exterior sources, stimulating wave processes. These processes are described by the four-dimensional Dalamber equation.

Electrodynamics also comes to four-dimensional wave equation. However, its derivation from Maxwell's equations uses Coulomb and Lorenz gauge conditions. But there are no similar conditions in continuum mechanics. Thus, the formal coincidence of mechanics and electrodynamics occurs when writing wave equations, and while dealing with individual phenomena analogies are not always present.

**Theoretical analysis**

Let us use formal field approach in the construction of electrodynamics. As field phenomena are considered, four-dimensional Dalamber equation should be used to describe them. In essence, this is the postulate of general field theory. Naturally, a question about physical meaning of the four-dimensional function, which is the main characteristic of the electromagnetic field, arises. In macroscopic theory it is a four-dimensional vector potential $(\vec{A}, \varphi)$, so it is possible to write down well-known equations:

$$\Delta \vec{A} - \varepsilon\mu \frac{\partial^2 \vec{A}}{\partial t^2} = -\mu \vec{j}, \qquad (1)$$

$$\Delta \varphi - \mu\varepsilon \frac{\partial^2 \varphi}{\partial t^2} = -\frac{\rho}{\varepsilon}, \qquad (2)$$

where $\varepsilon$ and $\mu$ - respectively, dielectric and magnetic permeability.

As the cause and effect are always separated spatially and in time, coordinate-time continuum field sources and characteristics of the field itself must be distinguished. We characterize the domain of the currents and charges by coordinates with strokes and domain of field potentials – by coordinates without strokes. Equations' solutions (1) and (2) are written in the form of retarded potentials [2]:

$$\vec{A}(x,y,z,t) = \frac{\mu}{4\pi} \int_\tau \frac{\vec{j}(x',y',z',t-r/V)}{r} d\tau, \qquad (3)$$

$$\varphi(x,y,z,t) = \frac{1}{4\pi\varepsilon} \int_\tau \frac{\rho(x',y',z',t-r/V)}{r} d\tau, \qquad (4)$$

where $|\vec{r}| = \sqrt{(x'-x)^2 + (y'-y)^2 + (z'-z)^2}$ - is the distance between the source and the point of field determination, $V$- is absolute velocity of the wave process distribution, $\tau$ - volume of the area containing the sources.

Introducing four-dimensional space:
$$x_1 = x, \quad x_2 = y, \quad x_3 = z, \quad x_4 = ict.$$

Writing down the relevant components of the vector potential:
$$\Phi_1 = A_x, \quad \Phi_2 = A_y, \quad \Phi_3 = A_z, \quad \Phi_4 = ic\varphi,$$

and four-dimensional wave equation, consolidating (1) и (2):

$$\Box \Phi_\nu = -\mu s_\nu. \qquad (5)$$

Here $\Box$ - is invariant of Dalamber operator, and components of the four-dimensional vector current density are used

$$s_1 = \rho V_x, \quad s_2 = \rho V_y, \quad s_3 = \rho V_3, \quad s_4 = ic\rho.$$

Calculating four-dimensional divergence of the vector potential:

$$\frac{\partial \Phi_1}{\partial x_1} + \frac{\partial \Phi_2}{\partial x_2} + \frac{\partial \Phi_3}{\partial x_3} + \frac{\partial \Phi_4}{\partial x_4} = div\vec{A} + \varepsilon\mu \frac{\partial \varphi}{\partial t}. \qquad (6)$$

Typically, from mathematical considerations, this expression is identically equal to zero, that is, Lorenz condition is applied. Let us investigate the issue of limited physical content of the theory in connection with this approach.

Usually particular (three dimensional) case of the Lorenz condition - Coulomb gauge, is used in magnetostatics:

$$div\vec{A} = 0,$$

eliminating potential component of the vector $\vec{A}$. Investigating magnetostatic solution:

$$\vec{A}(x,y,z) = \frac{\mu}{4\pi} \int_\tau \frac{\vec{j}(x',y',z')}{r} d\tau. \qquad (7)$$

We know that after applying operator *rot* to it, the result is vortex characteristic of the field $\vec{B}(x,y,z)$. Applying the operator div to (7), we obtain:

$$div\vec{A}(x,y,z) = \frac{\mu}{4\pi} \int_\tau div \frac{\vec{j}(x',y',z')}{r} d\tau.$$

Here we have that the order of integration and calculating the divergence on the right side can be changed, since they are executed on different coordinates. Transform the integrand:

$$div \frac{\vec{j}(x',y',z')}{r} = \frac{1}{r} div \vec{j} + \vec{j} \cdot grad \frac{1}{r} = -\frac{\vec{j} \cdot \vec{r}}{r^3}.$$

Here $div\vec{j}(x',y',z') = 0$, because when calculating *div*, differentiation is on unprimed coordinates. Since the result is nonzero, we introduce a scalar function $B^*(x,y,z)$, and write:

$$B^*(x,y,z) = \frac{\mu}{4\pi} \int_\tau \frac{\vec{j} \cdot \vec{r}}{r^3} d\tau. \tag{8}$$

That is, an analog of the Biot-Savart law [3-4]. Scalar function $B^*(x,y,z)$ characterizes potential component of the magnetic field, which is usually excluded from the application by Coulomb gauge and, therefore, it is virtually unexplored. Its physical richness of content is proved theoretically and experimentally in publications [3-4]. Conditions of the potential magnetic field appearing are determined. In particular, it is shown that on the current's section, placed in the potential magnetic field, a force directed along the current or against it appears, depending on the sign of $B^*(x,y,z)$.

It follows, that for a complete description of the magnetic field it is required to use a four-dimensional vector $(\vec{B}, B^*)$, which reflects its potential-vortex character. Thus, rejection of the Coulomb gauge, allows to take into account both potential and vortex components of the vector $\vec{A}$. General field theory also suppose this. Let us formulate the fundamental theorem of Stokes-Helmholtz [5] for vector $\vec{A}$: if field's divergence and rotor vanish at infinity, defined at each point $\vec{r}$ of a certain area, then vector's $\vec{A}$ field can be uniquely represented everywhere in this area (up to a constant vector) as a sum of potential and solenoidal fields:

$$\vec{A} = \vec{A}_\rightarrow + \vec{A}_\circ.$$

That is, the Stokes-Helmholtz theorem also requires to exclude Coulomb gauge and introduce scalar function in the stationary case:

$$B^*(x,y,z) = -div\vec{A} = -div\vec{A}_\rightarrow. \tag{9}$$

Obviously, the result of evaluating the four-dimensional divergence of the vector potential $(\vec{A},\varphi)$ is also not identical to zero. That is, in the non-stationary case, instead of the Lorentz condition, the following should be written:

$$B^*(x,y,z,t) = -div\vec{A} - \varepsilon\frac{\partial\varphi}{\partial t}. \tag{10}$$

Usually in the educational literature proof of Stokes- Helmholtz theorem is given for the stationary case. But it can be easily generalized for the non-stationary case. This approach is usually attached to the non-stationary electric field, vector $\vec{E}(\vec{r},t)$ of which fully satisfies the Stokes-Helmholtz theorem. However, application of this theorem to the magnetic field is a problem, because it does not have the potential component in classical electrodynamics. Proposed higher field approach solves this problem, since both components of the magnetic field are taken into account: vortex $\vec{B}(\vec{r},t)$ and potential $B^*(\vec{r},t)$.

**Properties of vector electrodynamic potential**

Stokes-Helmholtz theorem usually is not applied directly to the vector $\vec{A}$ because it is considered a support function and is ambiguously determined. Let us examine the properties of the vector $\vec{A}$ [3-4]. In classical electrodynamics the following correlations are used:

$$\vec{B} = rot\vec{A}, \quad \vec{E} = -\frac{\partial\vec{A}}{\partial t} - grad\varphi. \tag{11}$$

They are usually used gradient transformation:

$$\vec{A}' = \vec{A} + grad\chi, \quad \varphi' = \varphi - \frac{\partial\chi}{\partial t}, \tag{12}$$

where $\chi$ - arbitrary scalar coordinate-time function. Here $\vec{B}$, $\vec{H}$ characteristics of the vortex magnetic field, and $\vec{E}_O$, $\vec{D}_O$ characteristics the vortex electric field are invariant under the transformation (12). On this basis, conclusion about gradient invariance of the electromagnetic field is made. This conclusion is the basis for the introduction of the Coulomb gauge and the Lorentz condition. No physical meaning is normally recognized in the transformation (12).

Let us try to explain physical meaning of gradient transformation. Note that by adding $grad\chi$ to the vector potential, its potential part changes. Potential part of a vector field changing without vortex component's changing is possible only in the transition from conventional fixed reference frame $K$ to a steadily moving reference frame $K'_n$. But in this case, obviously, potential part of the electric field in the direction of the reference system must change. In moving reference frame the electric field of a point charge is not spherically symmetric, but appears Heaviside ellipsoid [6]. To compensate this change ("deformation"), an additive $\frac{\partial \chi}{\partial t}$ with a minus sign is introduced to the second correlation. Does the scalar potential $\varphi$ really change at the transition from $K$ to $K'_n$? It is known [6] that the charge is relativistically invariant value. Scalar potential $\varphi$ depends on the location of the point of its definition and the magnitude of the charge, therefore, there is reason to consider it in all reference frames the same (at speeds $K'_n$ much smaller than the speed of light). "Deformation" of the electric field in the transition to the mobile reference system is fully taken into account by vector potential $\vec{A}$. That is, there is no need to introduce a second correlation (12), it has no physical meaning. To describe the transformation of the electric field in the transition between $K$ and $K'_n$ (gradient transformation) only the first relation (12) is necessary.

Physical considerations suggest two possible types of transformations of the vector $\vec{A}$ field: gradient and vortex. Gradient transformation, as has been shown above, corresponds to the transition between progressively moving reference frames (one of them can be considered relatively immobile). When vortex transform, a transition from progressively moving (or conditionally stationary) reference frame $K$ to the – to the rotating $K'_{вр}$.

It can be shown rigorously [3-4] that at the gradient transformation potential characteristics of the electromagnetic field change ($\vec{A}_\rightarrow$, and, consequently, $\vec{E}_\rightarrow$ and $B^*$), vortex ($\vec{A}_O$, and, consequently $\vec{E}_O$ and $\vec{B}$) - are invariants. When vortex transformation, on the contrary, vortex components of the electromagnetic field change, and potential - are invariants. This corresponds to the relative nature (depending on the choice of the reference frame) of the magnetic field and its basic characteristics - vector $\vec{A}$.

Let us consider the case when particle moves along the vortex line of the vector $\vec{A}$ field. Natural trihedral can associated with it, which performs a complex motion consisting of translational and rotational components in relation to the laboratory system. In this particular case vortex component of the vector in the accompanying particle reference frame is completely absent, the field seems purely potential. We can say that as a result of gradient-vortex transformation, transformation of the vortex field to the potential occurs, and vice versa.

So, to conclude: the correlation of solenoidal and potential components of the electrodynamic vector $\vec{A}$ depends on the choice of the reference system, but in the selected reference frame is uniquely determined. Information contained in brackets in the Stokes-Helmholtz theorem, merely reflects the relative character of motion (rest) of any reference system. In selected arbitrarily fixed reference frame it is convenient this vector constant equal to zero. In this approach, the ambiguity of the choice of potentials $\vec{A}$ and $\varphi$ disappears, and there is no need to introduce gauge conditions. A theory formed on this platform will be called generalized electrodynamics.

**Modified (generalized) Maxwell equations**

From the wave equations (1) and (2) taking into account correlations (10) and (11), it is easy to obtain equations of generalized electrodynamics (modified Maxwell's equations) [3-4]:

$$rot\vec{H} + gradH^* = \vec{j} + \frac{\partial \vec{D}}{\partial t}, \qquad (13)$$

$$div\vec{D} = \rho + \varepsilon'\varepsilon_0 \frac{\partial B^*}{\partial t}. \qquad (14)$$

General field approach is not possible to exclude the potential component of the magnetic field in these equations, which is described by two interrelated scalar functions:

$$B^* = \mu H^*. \qquad (15)$$

The second term, as contained in the right side of equation (14), describes the phenomenon of vortex-free electromagnetic induction, confirmed experimentally [3-4].

Two more differential equations are also associated with already used correlations (11):

$$rot\vec{E} = -\frac{\partial \vec{B}}{\partial t}, \qquad (16)$$

$$div\vec{B} = 0. \qquad (17)$$

Note that the right sides of equations (13) - (14), except for the above-mentioned field sources $\vec{j}$ and $\rho$ contain expressions describing non-stationary electromagnetic processes. The right side of equation (16) also describes a non-stationary process. In Maxwell's electrodynamics only non-stationary vortex processes are considered. Generalized electrodynamics describes, in addition, and non-stationary processes that give rise to potential electric and magnetic fields. If these non-stationary processes are created by exterior generators, they are the real sources of the electromagnetic field. In the case of free field they should be regarded as quasi-sources.

Equations (13) - (14), (16) - (17) and their functions describe some separate electromagnetic phenomena, and for a complete description of the electrodynamic process, wave equations (1) - (2) and basic electrodynamic characteristics - four-dimensional vector potential $(\vec{A}, \varphi)$ must be used. When attempting to describe wave electrodynamic process using Maxwell's equations (or equations of generalized electrodynamics) problems of substantial matter appear. If you describe the picture of the free electromagnetic wave difusion through the equations of electro-dynamics (without real sources), then at each stage quasi-source must be determined (ie the cause), tell the difference between its coordinates and coordinates of the characteristics of the field generated by it (ie the consequence), taking into account a delay. This procedure is difficult to describe and tedious to read. There is no need in it if the description of the wave process is based on the homogeneous differential equations corresponding to equations (1) - (2) or (5). Their solution allows us to determine all the characteristics of the electrodynamic process.

**Wave processes**

It is possible to make direct transition from (1) - (2) to the wave equations for the individual characteristics of the field. It is not difficult to obtain the wave equation from equations (1) - (2) using the second relation (11):

$$\Delta \vec{E} - \mu\varepsilon \frac{\partial^2 \vec{E}}{\partial t^2} = \mu \frac{\partial \vec{j}}{\partial t} + \frac{1}{\varepsilon} grad\rho . \qquad (18)$$

Conduction currents should be divided into secluded $\vec{j}_O$ and open-ended $\vec{j}_\rightarrow$. An example of an open-ended conduction current is the discharge between two divorced in the space charges. With this in mind, perhaps splitting (18) into two independent equations:

$$\Delta \vec{E}_\rightarrow - \mu\varepsilon \frac{\partial^2 \vec{E}_\rightarrow}{\partial t^2} = \mu \frac{\partial \vec{j}_\rightarrow}{\partial t} + \frac{1}{\varepsilon} grad\rho , \qquad (19)$$

$$\Delta \vec{E}_O - \mu\varepsilon \frac{\partial^2 \vec{E}_O}{\partial t^2} = \mu \frac{\partial \vec{j}_O}{\partial t} . \qquad (20)$$

Using the first correlation (11) from (1) - (2) the wave equation for the the vector $\vec{H}$ is obtained:

$$\Delta \vec{H} - \varepsilon\mu \frac{\partial^2 \vec{H}}{\partial t^2} = -rot\vec{j}_O . \qquad (21)$$

Well-known mechanism of radiation of transverse electromagnetic waves is explained on the basis of equations (20) and (21).

Similarly, transforming (1) - (2) with (10), the wave equation for the scalar function $H^*$ is obtained:

$$\Delta H^* - \varepsilon\mu \frac{\partial^2 H^*}{\partial t^2} = \frac{\partial \rho}{\partial t} + div\vec{j}_\rightarrow . \qquad (22)$$

Differential Equations (19) and (22) together describe the process of wave distribution along not whirling vector, so they should be called longitudinal [2-3] or electro-scalar [8].

It is important to note that equations (19), - (22) are fully consistent with the superposition principle: that is, vortex and potential components of the fields are mutually independent. The sources generating potential and vortex fields are separated, respectively. However, the question arises about the correlation between the unconfined conduction currents and charges, which the condition of continuity reflects. In Maxwell's electrodynamics, the condition of continuity is used in the following form:

$$\frac{\partial \rho}{\partial t} + div \vec{j}_\rightarrow = 0. \tag{23}$$

In the generalized theory, from differential equations (14), it follows that the formation of an electric current in a conductive medium is possible not only due to changes in electrical charges to a certain extent, but also because of non-stationary process, defined by the change of a scalar function $B^*$. In this approach, more general condition of continuity [3-4]:

$$\frac{\partial \rho}{\partial t} + \varepsilon' \varepsilon_0 \frac{\partial^2 B^*}{\partial t^2} + div \vec{j}_\rightarrow = 0. \tag{24}$$

It contains a term depending on the induction of the non-stationary potential magnetic field. Consequently, the current can be created without the aid of non-stationary charges, but due to the alternating magnetic field of potential (vortex-free electromagnetic induction [3-4]). In this sense, the usual electric charges and currents can be independent. Incidentally, Zhilin P.A. comes to the same conclusion [1], based on the general provisions of the field theory.

Using (24), equation (22) takes form:

$$\Delta H^*(x,y,z,t) - \varepsilon\mu \frac{\partial^2 H^*(x,y,z,t)}{\partial t^2} = -\varepsilon \frac{\partial^2 B^*(x',y',z',t-r/V)}{\partial t^2}. \tag{25}$$

Non-stationary process defined by the right side, in this case corresponds to a real exterior source (not quasi-source), in addition that is created by an external generator. Values in (25) on the right (the reason) and left (result) belong to different points of space-time continuum, so the second term of the left side is not compensated by a similar right-wing member.

The question of the technical unit of the potential magnetic field generator is extremely interesting and has great practical significance. Creation of such a generator will allow transmission of electromagnetic signals using longitudinal (electroscalar) electromagnetic waves. The first successful experiments are described in [7].

In the field approach it becomes apparent that Maxwell's equations (or generalized equations of electrodynamics) play only a supportive role and are only suitable for describing some of electrodynamic phenomena. Their integration into the mathematical system does not lead to physical integration of disparate electrodynamic phenomena. This occurs only at the level of the wave differential equations (1) - (2). Use Maxwell's equations (or generalized equations of electrodynamics) to describe the wave process makes no sense, since more clearly, this picture seems to equations (1) - (2) or (5), as well as equations (19) - (22).

**Conclusion**

From the above it is possible to conclude: the state and evolution of the electro-magnetic field in a macroscopic approximation to the selected reference frame is clearly described by four-dimensional vector, which includes potential and solenoidal components and satisfies four-dimensional Dalamber equation. No problems with system of equations' certainty at such approach arise. If we divide mutually independent potential and vortex parts of the vector $\vec{A}$, then from (1) - (2) seven scalar differential equations and seven independent variables for determination are obtained. All other characteristics of electromagnetic fields (intensity, induction) are secondary, they are uniquely expressed through the potential $(\vec{A},\varphi)$.

A very similar position expressed K. J. van Vlaenderen [9] in his article. The author of this article also pointed to the gauge conditions as the reason for the limitation of the electrodynamic theory and received modified equations that coincide with (13) - (14), (16) - (17). However, he did not use the continuity equation in a generalized form (24) and did not give up gauge invariance in its traditional interpretation. Such a piecemeal approach contains internal contradictions. Therefore, Bruhn G. W. [11] expressed doubts about the validity of the theory, developed by K. J van Vlaenderen.

Higher generalization of electrodynamics is the quantum level. It uses two four-dimensional vectors. Generalized equations of quantum electrodynamics led Hvorostenko N.P. [10], their analysis is contained in papers [3-4]. A similar result for quantum processes was received by Dale A. Woodside [12].

Using a generalized electromagnetic theory opens up new opportunities for engineering and technology. First of all, it concerns radio physics and electricity, as well as methods of recording and storing information.

**Литература**
1. Zilin P.A.   http://www.spbstu.ru/public/m_v/lib/Zhilin/RealyTim.PDF
2. Batigin V.V.,   Toptigin I.N.   http://www.twirpx.com/file/35115/
3. Tomilin A. K.   http://www.spbstu.ru/public/m_v/N_017/Tomilin_01.pdf
4. Tomilin A. K.  The Fundamentals of Generalized Electrodynamics. http://arxiv.org/pdf/0807.2172
5. Helmholtz H. Uber Integrale der hydrodynamischen Gleichungen, welche den Wirbelbewegungen entsprechen. Crelles J. 55, 25 (1858). http://puhep1.princeton.edu/~mcdonald/examples/fluids/helmholtz_jram_55_25_58.pdf
6. Purcell E. Electricity and Magnetism. Berkeley Physics Course. V. 2. McGraw-. Hill, New York, 1963 – 430 p.
7. Monstein C., Wesley J. P.  Europhysics  Letters, 59 (4), pp. 514-520 (2002).

8. Arbab I. Arbab, Zeinab A. Satti. On the Generalized Maxwell Equations and Their Prediction of Electroscalar Wave// Progress in physics, 2009, v.2.- s. 8-13.

9. Koen J. van Vlaenderen. A generalisation of classical electrodynamics for the prediction of scalar field effects. http://arxiv.org/abs/physics/0305098v1

10. Bruhn G. W. On Koen van Vlaenderen's Seventh Field Component Again http://www.mathematik.tu-darmstadt.de/~bruhn/vV191005.html

11. Hvorostenko N.P. http://sto68.narod.ru/hvorostenko-24-25.jpg

12. Woodside D.A. Three-vector and scalar field identities and uniqueness theorems in Euclidean and Minkowski spaces// Am. J. Phys., Vol.77, № 5, pp.438- 446, May 2009. woo09ajp.pdf